# A Lightweight Domain Adversarial Neural Network Based on Knowledge Distillation for EEG-based Cross-subject Emotion Recognition

Zhe Wang, Yongxiong Wang, Jiapeng Zhang, Yiheng Tang, Zhiqun Pan

***Abstract*—Individual differences of Electroencephalogram (EEG) could cause the domain shift which would significantly degrade the performance of cross-subject strategy. The domain adversarial neural networks (DANN), where the classification loss and domain loss jointly update the parameters of feature extractor, are adopted to deal with the domain shift. However, limited EEG data quantity and strong individual difference are challenges for the DANN with cumbersome feature extractor. In this work, we propose knowledge distillation (KD) based lightweight DANN to enhance cross-subject EEG-based emotion recognition. Specifically, the teacher model with strong context learning ability is utilized to learn complex temporal dynamics and spatial correlations of EEG, and robust lightweight student model is guided by the teacher model to learn more difficult domain-invariant features. The proposed method contains two stages, and temporal-spatial feature interaction is adopted throughout the two stages. In the feature-based KD framework, a transformer-based hierarchical temporal-spatial learning model is served as the teacher model. The student model, which is composed of Bi-LSTM units, is a lightweight version of the teacher model. Hence, the student model could be supervised to mimic the robust feature representations of teacher model by leveraging complementary latent temporal features and spatial features. In the DANN-based cross-subject emotion recognition, we combine the obtained student model and a lightweight temporal-spatial feature interaction module as the feature extractor. And the feature aggregation is fed to the emotion classifier and domain classifier for domain-invariant feature learning. To verify the effectiveness of the proposed method, we conduct the subject-independent experiments on the public dataset DEAP with arousal and valence classification. The outstanding performance and t-SNE visualization of latent features verify the advantage and effectiveness of the proposed method.***

## I. Introduction

EMOTION recognition has received more attention and become the hotspot of neuroscience and computer science in recent years. Emotion recognition aims to build a harmony human-computer interface by endowing the computer to recognize and comprehend the emotions, so it has great significance to the commerce and scientific research. Many modalities could be adopted to represent the emotions, such as images [1], texts [2], and physiological signals [3]. Electroencephalogram (EEG) as a typical physiological signal is highly associated with emotions [4]. However, strong individual differences and limited data quantity [5] would decrease the performance of cross-subject emotion classification. Hence, it has become the challenge to the application of affective brain computer interface (aBCI) in real life.

Studies on EEG-based emotion recognition are commonly based on the independent identically distributed (IID) hypothesis. It means that samples of training set and testing set are assumed to be distributed in the same domain. However, the performance of these methods would significantly decrease in the cross-subject emotion recognition, because the samples of training and testing set are from different subjects and the feature distributions among EEG subjects are inconsistent.

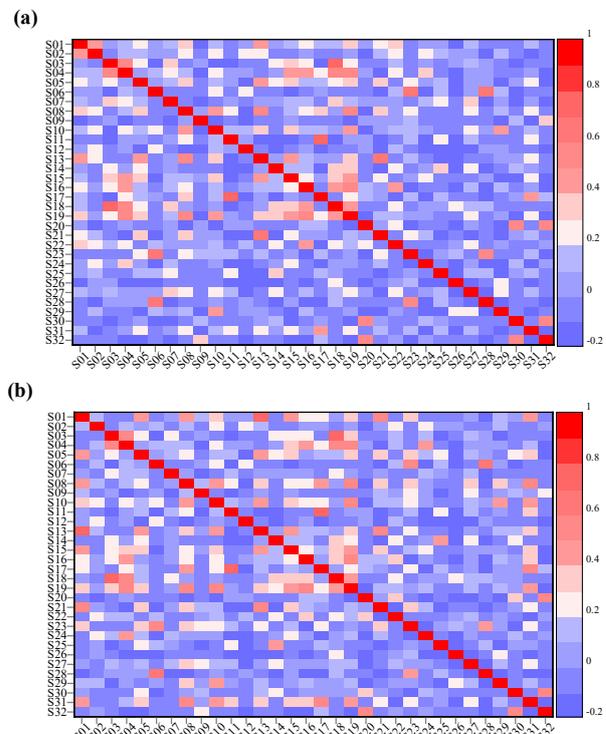

Fig. 1. Feature correlation analysis based on Pearson correlation analysis. (a) PSD feature correlations on beta band (12-30 Hz) (b) PSD feature correlations on gamma band (30-45 Hz)

This work is sponsored by Natural Science Foundation of Shanghai under Grant No. 22ZR1443700. *(Corresponding author: Yongxiong Wang)*

Z. Wang, Y. Wang, J. Zhang, Y. Tang, and Z. Pan are with the School of Optical-Electrical and Computer Engineering, University of Shanghai for Science and Technology, Shanghai 200093, China (e-mail: 201440049@st.usst.edu.cn; wyxiong@usst.edu.cn; mnrsmzjp@126.com; 213330685@st.usst.edu.cn; panzhiqun@139.com).

In the DEAP dataset, the correlations of power spectral density (PSD) feature among all the subjects on beta and gamma band are shown in Fig.1. The white color is according to the correlation coefficient with 0.2-0.3. Hence, the feature correlation between two subjects increases gradually from white to red. We find that small part of the cases have exceeded 0.4, and about half of the cases have exceeded 0.2. These results could confirm the strong individual differences of EEG to some extent.

Domain adversarial neural networks (DANN) have been introduced to solve domain-shift problem. Tian et al. [6] propose a 3DLSTM-ConvNET based DANN to extract the discriminative features from temporal, frequency and spatial domains, and they adopt the local and global domain classifier to achieve domain-alignment. Li et al. [7] propose an asymmetric pattern inspired DANN to capture the spatial correlations of left and right hemispheres and reduce the domain differences by using multiple domain classifiers. He et al. [8] combine temporal convolutional networks (TCNs) and adversarial discriminative domain adaptation (ADDA) to solve the domain-shift problem. These methods have made great process in cross-subject emotion recognition. However, there is a potential problem in EEG-based DANN. EEG datasets commonly have limited data quantity problem due to the high cost of data sampling. And strong individual differences might be caused by subjects' sensory information to the stimuli materials [9], clinical settings and other reasons. If the feature extractor is cumbersome and complex, it would be difficult for the emotion classification loss and domain loss to update the weights of a large number of parameters due to the data quantity limitation and strong individual differences [10]. On the other hand, it is also difficult for simple feature extractor to extract discriminative features from complex temporal dynamics and spatial correlations of EEG. Hence, it is a challenge to reduce the size of the feature extractor within DANN meanwhile ensuring sufficient ability to extract discriminative temporal-spatial features.

Knowledge Distillation (KD) [11] is a method which assists the training of a smaller student network under the supervision of a larger teacher network. As a type of model compression method, a larger network can be downsized using KD method by minimizing the structural differences between the teacher and the student model. Feature-based KD and logit-based KD are the common choices. In the feature-based KD, richer information could be learned from the teacher model and it could provide more flexibility [12]. The goal of student model is to learn the feature representation through the corresponding robust intermediate representations of teacher model. And the similarity loss between the representations is valued via distance metrics. Finally, a robust lightweight student model could be obtained by inheriting the essential knowledge from teacher model.

In this work, we propose a two-stage method, where a robust lightweight feature extractor is firstly obtained by feature-based KD framework, to further improve the performance of DANN-based cross-subject emotion recognition. The teacher model is elaborately designed to robustly capture the temporal-spatial representations of EEG, and it is utilized to guide the lightweight student model. Then, the lightweight student model, which inherits the knowledge of robust feature representation, is more suitable as the feature extractor to learn the domain-invariant features. Furthermore, integrating complementary temporal and spatial information could provide more robust emotion prediction which is preliminarily explored in our previous research [13], so we appropriately design schemes to achieve the temporal-spatial feature interaction in the two stages. In the feature-based KD framework, the student model is supervised to approximate the robust and complementary temporal and spatial representations of teacher model. Specifically, the student model and pretrained teacher model are both end-to-end temporal-spatial hierarchical feature learning networks, and they are composed of Bi-LSTM units and transformer encoders [14] respectively. What's more, the teacher model is designed based on our previous work, hierarchical spatial learning transformer (HSLT) [15], and we propose a weight-shared temporal extractor to replace the handcraft feature extraction. In the DANN-based cross-subject emotion recognition, the temporal-spatial feature interaction network (TSFIN) is proposed, which combines the obtained student model and feature interaction module, to aggregate the latent temporal and spatial representations for domain-invariant learning. Moreover, the design of feature interaction module is inspired by multi-scale feature fusion in the Feature Fusion Single Shot Multibox Detector (FSSD) [16]. The contributions of this work can be concluded as follows:

(1) Inspired by the multi-representation adaptation network [17] which improve the performance by capturing the information from different aspects, we design a feature-based KD framework to obtain a robust lightweight model by leveraging comprehensive temporal and spatial representations. Specifically, the teacher model and student model are designed with similar structure, it ensures the transfer features with the same meaning and reduces the difficulty of following the teacher model [11,18]. Besides, the mainly difference between the transformer encoder and Bi-LSTM unit is the multi-head self-attention which has strong ability of sequence learning. The lightweight Bi-LSTM unit could inherit the knowledge from transformer encoder which contains a lot of parameters and consumes much more computation.

(2) Compared with other feature extractor within DANN, the proposed TSFIN as an FSSD-inspired network could provide comprehensive and complementary information for domain classifier and emotion classifier. Besides, Batch Normalization (BN) is a key component of deep semantic feature learning, but BN seems unfit to the sequence learning models [19], such as Transformer and Bi-LSTM. Hence, we add the BN in the feature interaction module to preserve the discriminative representation.

(3) We propose a weight-shared temporal feature extractor which is inspired by the handcraft feature extraction adopts the same rule among the EEG channels. And it could adaptively extract the essential information without domain knowledge.

The remainder of the paper is organized as follows. In the Section II, we review the related works. In Section III, we briefly introduce the proposed method. In Section IV, we firstly describe the DEAP dataset and preprocess the EEG data. And the experimental results of subject-independent experiments on the DEAP database are shown to validate the proposed method.

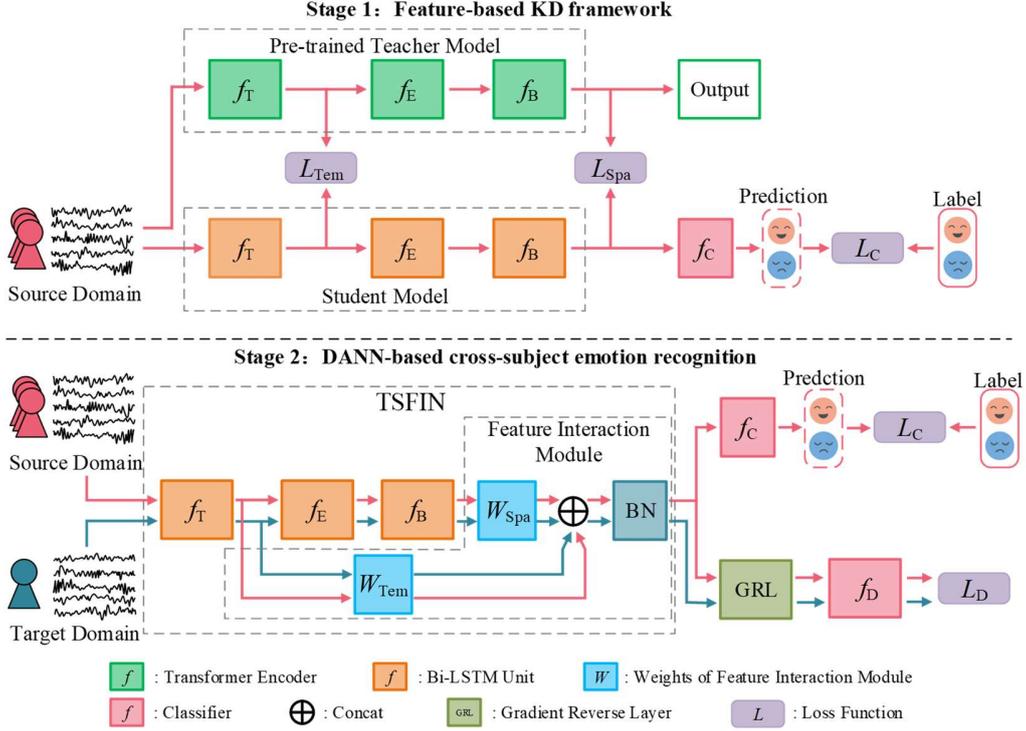

Fig. 2. The overview of the proposed method

Note: $f_T$ denotes weight-shared temporal feature extractor; $f_E$ and $f_B$ denote electrode-level spatial module and brain-region-level spatial module respectively; $f_C$ and $f_D$ denote emotion classifier and domain classifier respectively; $L_{Tem}$ and $L_{Spa}$ denote the temporal distillation loss and spatial distillation loss respectively; $L_C$ and $L_D$ denote the emotion classification loss and domain classification loss respectively; $W_{Tem}$ and $W_{Spa}$ denote the weights of feature interaction module.

In Section V, we discuss the visualization results and analyze the advantages and limitations of proposed method. Finally, we conclude the paper in Section VI.

## II. RELATED WORKS

### A. Spatial-temporal feature learning for EEG-based emotion recognition

High temporal-resolution and spatial correlations among the channels are the two major attributes of EEG. Hence, many researchers focus on the temporal and spatial EEG feature learning. Wang et al. [13] generate PSD based 32×32 spatial topographic maps and extract temporal dynamic features, the features are fed to a hybrid spatial-temporal fusion network. Ding et al. [20] propose a 1D-CNN based model which combines the spatial-temporal feature extraction and classification. Gao et al. [21] propose a Graph Convolutional Networks (GCN) to learn essential temporal-spatial information and capture the different activation pattern by using attention mechanism and adjacency matrix respectively. These researches indicate that the appropriate strategy of aggregating temporal and spatial information is beneficial to improve the performance.

### B. KD methods for different EEG tasks

In recent years, the KD methods have been achieved remarkable successes in different computer vison tasks, such as pedestrian detection [22], semantic segmentation [23], and depth estimation [25]. Nevertheless, few researchers adopt KD methods to improve the performance of different EEG tasks. In the emotion recognition, Wang et al. [25] propose the FLDNet to distill the essential frame-level representations net by net to replace the handcrafted features which would contain redundant information and loss frame-level context information. In the imagined speech recognition, García-Salinas et al. [26] propose a KD based incremental learning method to recognize new vocabulary of imagined speech while alleviating catastrophic forgetting problem. In the sleeping stage classification, Joshi et al. [27] propose a cross-modal KD framework to guide Electrocardiogram (ECG) feature learning by a pretrained EEG model. Different from these methods, we appropriately combine the KD and DANN to obtain a robust lightweight model for improving the performance of cross-subject emotion recognition.

## III. METHOD

### A. Overview

As shown in Fig.2, the proposed method contains two stages: the feature-based KD framework and DANN-based cross-subject emotion recognition. In the feature-based KD framework, the end-to-end teacher model is pretrained to assist the training of student model. Student model and teacher model both have the similar three parts: weight-shared temporal feature extractor electrode-level spatial learning module, and brain-region-level spatial learning module. And they are composed of Bi-LSTM units and transformer encoders respectively. Moreover, two distill losses and one classification loss are adopted to obtain a robust student model. Specifically, the corresponding outputs of temporal feature extractor and brain-region-level spatial learning module are adopted to compute two distill losses. In the DANN-based cross-subject emotion recognition, TSFIN, domain classifier, and emotion classifier are the main parts. What's more, the TSFIN includes the obtained student model and feature interaction module, where two adaptive weights are adopted to achieve temporal-spatial feature interaction, in order to obtain more discriminative representations. In the following of this section, we will briefly introduce the two stages.

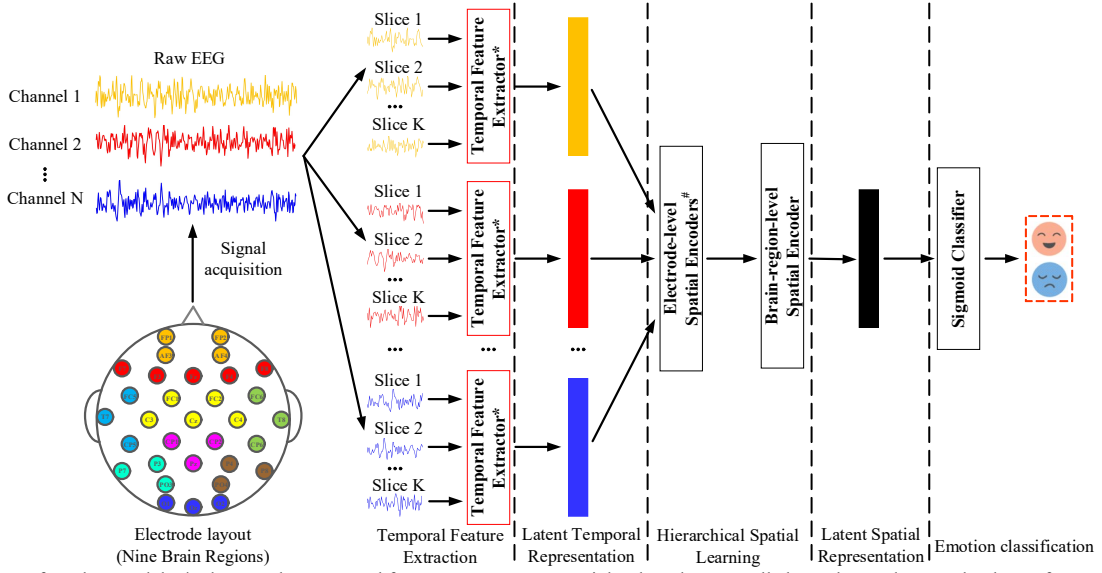

Fig. 3. The structure of teacher model. '*' denotes that temporal feature extractors are weight-shared among all channels. '#' denotes that latent features of different brain regions are separately fed to the corresponding electrode-level transformer encoders. And nine brain regions are classified according to different brain functions, the electrodes with the same color are divided into a brain region.

### B. Feature-based KD framework
#### 1. The teacher model

The proposed teacher model contains two crucial parts: temporal feature extraction and hierarchical spatial learning, as shown in Fig.3. Besides, the structure of student model is similar to the teacher model, and the transformer encoders within two parts are replaced by Bi-LSTM units. In the temporal feature extraction, raw EEG signals of each channel are equally divided into several slices. Next, EEG slices are considered as the input patches and fed into temporal feature extractors which share the weights among all the channels. Then the latent features could be obtained by the extractors, and they are divided into different feature sets according to the region classification of the cortex. In the hierarchical spatial learning, the feature sets are separately fed to the corresponding electrode-level transformer encoders. And the obtained representations of different brain regions are served as the patches to a brain-region-level spatial encoder in order to learn the global spatial information. Finally, a sigmoid classifier is utilized for the valence and arousal prediction.

Given representation of raw EEG signals with $N$ channels $X = [X^1, X^2, ..., X^N] \in \mathbb{R}^{N \times d}$, where $d$ is signal length. The weight of temporal feature extractor is share among all the channels. In each channel, the EEG signals are divided into $K$ patches $X^n = [X_1^n, X_2^n, ..., X_K^n] \in \mathbb{R}^{K \times d'}$ $n = 1,2, ..., N$, where is $d' = d/k$ is patch length. Firstly, electrode patches are mapped to a constant size $D_T$ using linear embedding with the weight $W_T \in \mathbb{R}^{d' \times D_T}$. Next, the class token $X_T^{cls} \in \mathbb{R}^{D_T}$ is inserted at the head of the patches to aggregate representative information. Besides, positional embedding $E_T^{pos} \in \mathbb{R}^{(K+1) \times D_T}$ is added to the patches to reserve positional information. For the $n_{th}$ electrode, these operations can be represented as follows:

$$Z_0^n = [X_T^{cls}, X_1^n W_T, ..., X_K^n W_T] + E_T^{pos} \quad (1)$$

where $Z_0^n \in \mathbb{R}^{(K+1) \times D_T}$ is the input of transformer encoder.

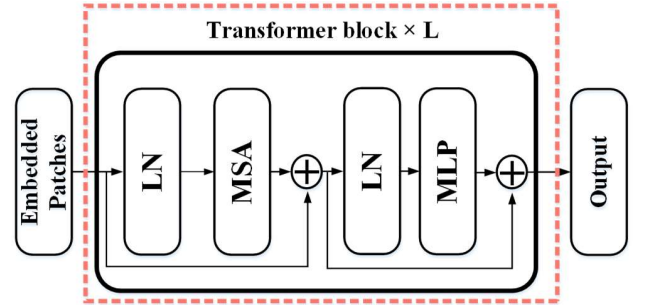

Fig. 4. The structure of transformer encoder

As shown in Fig.4, the transformer encoder includes: Multi-head Self-Attention (MSA), Multiple Layer Perception (MLP), and Layer Normalization (LN). Hence, the operation within the temporal transformer encoder is shown as:

$$\dot{Z}_l^n = G_{msa}(G_{ln}(Z_{l-1}^n)) + Z_{l-1}^n \quad l = 1,2, ..., L_T \quad (2)$$

$$Z_l^n = G_{mlp}(G_{ln}(\dot{Z}_l^n)) + \dot{Z}_l^n \quad l = 1,2, ..., L_T \quad (3)$$

where $G_{msa}(\cdot)$ denotes MSA operation, $G_{mlp}(\cdot)$ denotes MLP operation, and $G_{ln}(\cdot)$ denotes LN operation. Besides, the $L_T$ is the number of transformer blocks, and the final output is $Z_{L_T}^n \in \mathbb{R}^{(K+1) \times D_T}$. For each channel, we take mean of $(K + 1)$ patches as the latent feature $\tilde{Z}_{L_T}^n \in \mathbb{R}^{D_T}$. Finally, the obtained latent temporal features of all the channels are concatenated as $Z_{Tem} = [\tilde{Z}_{L_T}^1, \tilde{Z}_{L_T}^2, ..., \tilde{Z}_{L_T}^N] \in \mathbb{R}^{N \times D_T}$.

Before the spatial learning, the latent feature $Z_T$ are divided into nine subsets which has been preliminarily explored in our previous work [15]. In the electrode-level spatial learning, the operation of the $r_{th}$ brain region can be represented as follows:

$$Z_0^r = [X_E^{cls}, X_1^r W_E, ..., X_M^r W_E] + E_e^{pos} \quad r = 1, 2, ..., 9 \quad (4)$$
$$\dot{Z}_l^r = G_{msa}(G_{ln}(Z_{l-1}^r)) + Z_{l-1}^r \quad l = 1, 2, ..., L_E \quad (5)$$
$$Z_l^r = G_{mlp}(G_{ln}(\dot{Z}_l^r)) + \dot{Z}_l^r \quad l = 1, 2, ..., L_E \quad (6)$$

where $M \in \{3,4,5\}$ is the number of the electrodes within the brain region, $r$ is the index of the brain region, $Z_0^r \in \mathbb{R}^{(M+1) \times D_E}$ is the transformer input, $D_E$ is the patch dimension, $W_E \in \mathbb{R}^{D_T \times D_E}$ is the weight of linear embedding, $X_E^{cls} \in \mathbb{R}^{D_E}$ is the class token, and $E_e^{pos} \in \mathbb{R}^{(M+1) \times D_E}$ is the positional embedding. And $L_E$ is the number of transformer blocks, $Z_{L_E}^r \in \mathbb{R}^{(M+1) \times D_E}$ denotes the output patches. What's more, the average representation $\tilde{Z}_{L_E}^r \in \mathbb{R}^{D_T}$ of $Z_{L_E}^r$ is obtained for the next brain-region-level spatial learning.

Similar to the aforementioned, the operation within the brain-region-level spatial learning could be represented as:

$$Z_0^B = [X_B^{cls}, \tilde{Z}_{L_E}^1 W_B, ..., \tilde{Z}_{L_E}^9 W_B] + E_B^{pos}, Z_0^B \in \mathbb{R}^{(9+1) \times D_B} \quad (7)$$
$$\dot{Z}_l^B = G_{msa}(G_{ln}(Z_{l-1}^B)) + Z_{l-1}^B \quad l = 1, ..., L_B \quad (8)$$
$$Z_l^B = G_{mlp}(G_{ln}(\dot{Z}_l^B)) + \dot{Z}_l^B \quad l = 1, ..., L_B \quad (9)$$

The number of the block in the brain-region-level is $L_B$, the outputs $Z_{L_B} \in \mathbb{R}^{(9+1) \times D_B}$, and latent spatial representation $Z_{Spa} \in \mathbb{R}^{D_B}$ is the average of $Z_{L_B}$. Different from previous two steps, the output patch according to the class token $Z_{L_B}^0 \in \mathbb{R}^{D_B}$ is utilized to predict the emotion. The binary prediction of arousal or valence is obtained as follows:

$$\hat{y} = \sigma(W_O Z_{L_B}^0) \quad (10)$$

where $\hat{y}_t$ is the prediction of teacher model, $W_O$ is the weight of output layer, and $\sigma(\cdot)$ denotes the sigmoid function.

*2. The student model*

The structure of student model is similar to the teacher model, we replace the transformer encoders by Bi-LSTM units in the temporal feature extractors and hierarchical spatial learning. The Bi-LSTM network could capture the context information of the sequence by bidirectional learning [28], and it contains fewer parameters. Specifically, the temporal extractor of student model is a weight-shared Bi-LSTM network. For each electrode patch $X^n \in \mathbb{R}^{K \times d'}$, the according latent temporal representation $F_T^n$ could be obtained by:

$$F_T^n = G_t(X^n) \quad (11)$$
$$F_{Tem} = [F_T^1, F_T^2, ..., F_T^n] \quad (12)$$

where $G_t(\cdot)$ is the operation of temporal Bi-LSTM network, $F_T^n \in \mathbb{R}^{2d_t}$ is the output of the Bi-LSTM unit which combines the results of two direction, $d_t$ is the mapping dimension of the hidden layer, and $F_{Tem} \in \mathbb{R}^{N \times 2d_t}$ is the latent feature aggregation of all the EEG channels.

Similarly, we divided $L_{Tem}$ into nine subsets, the operation within $r_{th}$ brain region can be formulated as:

$$F_e^r = [G_e(F_r^1), G_e(F_r^2), ..., G_e(F_r^M)] \quad (13)$$

where $G_e(\cdot)$ is the operation of electrode-level spatial Bi-LSTM network, the output is $F_e^r \in \mathbb{R}^{2d_e}$, and $d_e$ is the dimension of the hidden layer. Finally, we obtained the prediction of student model $\hat{y}_s$ as follows:

$$F_{Spa} = [G_b(F_e^1), G_b(F_e^2), ..., G_b(F_e^9)] \quad (14)$$
$$\hat{y}_s = \sigma(W_O' F_{Spa}) = G_c(F_{Spa}) \quad (15)$$

where $G_b(\cdot)$ is the operation of brain-region-level spatial Bi-LSTM network, $F_{Spa} \in \mathbb{R}^{2d_b}$ is the latent spatial features, $d_b$ is the number of neurons in the hidden layer, $W_O'$ is the weight of output layer, and $G_c(\cdot)$ is the classifier function.

*3. The training of feature-based KD framework*

Feature-based distillation enables learning richer information from the teacher and provides more flexibility for performance improvement. In this work, we apply the latent temporal and spatial representations within the hierarchical network to achieve feature-based KD. Before the KD operation, the dimension of the latent features from the student model should be consistent with the features from teacher model.

$$F_{Tem}' = F_{Tem} W_{Tem} \quad (16)$$
$$F_{Spa}' = F_{Spa} W_{Spa} \quad (17)$$

where the $F_{Tem}' \in \mathbb{R}^{N \times D_T}$ and $F_{Spa}' \in \mathbb{R}^{D_B}$ are the converted representations, $W_{Tem}$ and $W_{Spa}$ are the transformation matrices. And we adopt the mean squared error (MSE) loss to align the temporal and spatial representations from the teacher and student model. These operations could be formulated as:

$$L_{Tem} = \frac{1}{n_s} \sum_{X \in D_S} L_{mse}(F_{Tem}', Z_{Tem}) \quad (18)$$
$$L_{spa} = \frac{1}{n_s} \sum_{X \in D_S} L_{mse}(F_{Spa}', Z_{Spa}) \quad (19)$$

where $L_{Tem}$ and $L_{spa}$ are the distillation losses, $n_s$ is the sample number of the source domain, $D_s$ denotes the source domain, $L_{mse}(\cdot)$ is the MSE loss. Finally, the distillation loss and classification loss are jointly utilized to train student model.

$$L_{stu} = \frac{1}{n_s} \sum_{X \in D_s} L_c\left(G_c\left(G_b\left(G_e(G_t(X))\right)\right), y\right)$$
$$+ \lambda * L_{Tem} + (1 - \lambda) * L_{Spa} \quad (20)$$

where $L_{stu}$ is the total training loss of the student model, $L_c(\cdot)$ is the classification loss, $y$ is the emotion label, and $\lambda$ is a trade-off hyper-parameter.

*B. DANN-based cross-subject emotion recognition*

Feature extractor, domain classifier and emotion classifier are the mainly three parts of DANN-based cross-subject EEG emotion recognition. Here, we firstly introduce feature extractor, TSFIN. The TSFIN contains two parts: the obtained student model and FSSD-inspired feature interaction module. Compared with prevalent CNN, the multi-scale features from different intermediate layers are aggregated in the FSSD which aims to obtain abundant semantic information. Inspired by this, we design the feature interaction module to aggregate complementary temporal and spatial representation. Operation within the feature fusion module $G_{fusion}(\cdot)$ can be formulated as:

$$\dot{F}_{Tem} = G_{l-relu}(G_{mlp}(F_{Tem})) \quad (21)$$
$$\dot{F}_{Spa} = G_{l-relu}(G_{mlp}(F_{Spa})) \quad (22)$$

$$F_{Fusion} = G_{bn}(\dot{F}_{Tem} \cup \dot{F}_{Spa}) \tag{23}$$

where $G_{l-rel}(\cdot)$ denotes the leaky ReLU function, $\dot{F}_{Tem}$ and $\dot{F}_{spa}$ are concatenated to obtain the temporal-spatial feature aggregation $F_{Fusion}$ which is fed to the domain classifier and emotion classifier. Specifically, sequence learning models, such as Transformer and Bi-LSTM, focus on the intra-sample context representation, so inter-sample BN is not commonly utilized. However, BN could significantly improve the performance of classification task. Hence, we adopt the $G_{bn}(\cdot)$ in the feature fusion module. Finally, the function of feature extractor $G_f(\cdot)$ is equal to $G_{fusion}(G_b(G_e(G_t(\cdot))))$.

DANNs are designed based on GAN [29] to achieve domain-alignment. The core of DA adaption is that the feature extractor attempts to make domain discriminator confused about the samples from the source domain and target domain by extracting domain-invariant features. The training objective of DA network is minimizing the emotion classification loss $L_c$ while maximizing the domain classification loss $L_d$. The loss function can be formalized as:

$$L(\theta_f, \theta_c, \theta_d) = \frac{1}{n_s} \sum_{X \in D_s} \left( G'_c\left(G_f(X)\right), y \right)$$
$$- \frac{1}{n_s + n_t} \sum_{X \in (D_s \cup D_t)} (G_d(G_f(X)), d) \tag{24}$$

where $L$ is the total loss, $\theta_f, \theta_c, \theta_d$ are the parameters of feature extractor, emotion classifier $G'_c(\cdot)$ and domain classifier $G_d(\cdot)$ respectively. And $L_d$ is the domain classification loss, $n_t$ is the number of target domain sample, and $D_t$ denotes the target domain, and $d \in \{0,1\}$ is the domain label, where 0 denotes the source domain and 1 denotes the target domain. Although the optimization of the emotion classification loss and domain classification loss is reversed, the parameters can be trained efficiently with the Gradient Reversal Layer (GRL). When the network converges, the parameters $\theta_f, \theta_c, \theta_d$ will reach a saddle point as follows:

$$(\hat{\theta}_f, \hat{\theta}_c) = arg\ min_{\theta_f, \theta_c} L(\theta_f, \theta_c, \theta_d) \tag{25}$$
$$(\hat{\theta}_d) = arg\ max_{\theta_d} L(\theta_f, \theta_c, \theta_d) \tag{26}$$

## IV. EXPERIMENTAL RESULTS

### A. Data and experimental settings

In this work, the DEAP database [30] is chosen as the benchmark emotion databases. The DEAP is a multimodal emotion database which includes EEG signals and peripheral signals from 32 subjects. The 40 one-minute-long music videos are utilized to elicit the emotions, and these videos are presented in 40 trails. In each trail, the EEG signals and peripheral physiological signals are simultaneously recorded when subjects watching the videos. At the end of each trail, the subjects are rated the arousal, valence, and dominance according to Self-Assessment Manikin (SAM) with a range of 1 to 9.

The preprocessing of EEG contains three steps. Firstly, the 32-channel EEG signals are down-sampled from 512Hz to 128Hz. Next, a 4-45 Hz band pass filter and independent component analysis are adopted to remove the artifacts. Besides, the stimulated trial signals of each EEG channel subtract the average of baseline signals from corresponding channel. Finally, we adopt a 6-second-long sliding window without overlap to segment the EEG data. Each segment within a trial is considered as a sample.

On the other hand, the self-assessment value of 1-4 is considered as 'low class' the low arousal (LA) and low valence (LV). And the self-assessment value of 6-9 is considered as 'high class' the high arousal (HA) and high valence (HV). Therefore, we transform the emotion recognition problem into two binary classifications (LA vs. HA, and LV vs. HV).

In this section, we conduct the subject-independent experiments on the DEAP database for arousal and valence classification to validate the effectiveness proposed method. In each subject-independent experiment, the leave-one-subject-out cross-validation are adopted to derive the classification performance. Specifically, EEG data from one subject is considered as the testing set (target domain), and EEG data from the remaining subjects are set as training set (source domain) until each subject's data has been set as test set once. And the final performance is the average of all the folds.

The hyperparameters are also necessary to determine. In the teacher model, we set the $d$ as 768 (6s × 128Hz), $K$ as 6 and $d'$ as 128. In the module of temporal feature extraction, the $D_T$ and $L_T$ are set as 64 and 1 respectively. In the module of hierarchical spatial learning the $D_E$, $D_B$, $L_E$, and $L_B$ are set as 32, 64, 2 and 2 respectively. In the student model, we set $d_t$, $d_e$, $d_b$ as 64, 32, 32 respectively. In the KD stage, the trade-off parameter $\lambda$ is set as 0.3. Compared with latent temporal features, we give a bigger weight for the deep latent spatial features. We adopt Adam optimizer with learning rate is $10^{-3}$, batch size is 512 and epoch is 60 with early stopping. In the DANN stage, we adopt the Adam optimizer with learning rate is $10^{-3}$, batch size is 256 and epoch is 80 with early stopping. In this work all the networks are implemented by Pytorch with a NVIDIA GeForce RTX 3090 GPU.

### B. The experimental results of the proposed method

The overall results of proposed method in the subject-independent experiment are shown in Table I and detailed results of all subject are shown in Fig.5. We adopt the accuracy (denoted as $P_{acc}$) and F1-score (denoted as $P_f$) to evaluate the classification performance. In the Table.1, we utilize two strategies, all subjects' performances and top-10 subjects' performances, to evaluate the proposed method.

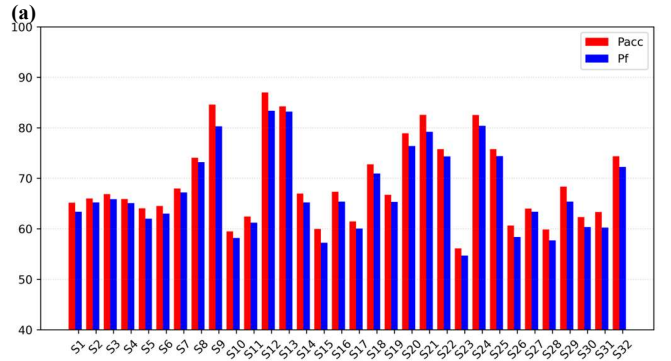
(a)

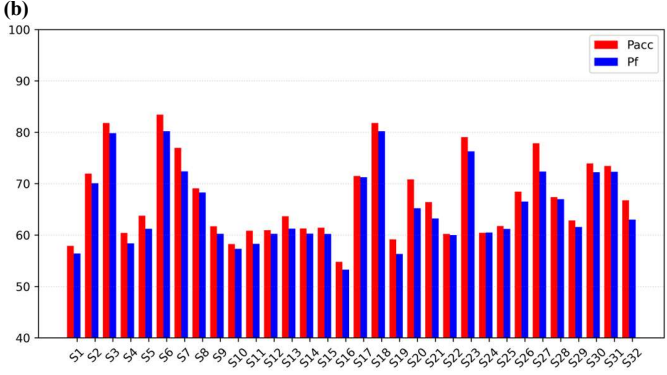
Fig. 5. The detailed results of all subjects (a) Arousal (b) Valence

In the arousal classification, the $P_{acc}$ of all subjects achieves 68.13±8.45%, and $P_f$ is 66.61±8.53%. According to the Fig.5., S12, S9, S13, S21, S24, S20, S22, S32, S8 and S18 are the subjects who achieves the top-10 accuracy. The $P_{acc}$ of top-10 subjects achieves 78.68±4.88%, and the corresponding $P_f$ is 77.16±4.68%. On the other hand, the $P_{acc}$ of all subjects in the valence classification achieves 67.03±8.08%, and $P_f$ is 65.87±7.87%. Meanwhile, S6, S3, S18, S23, S27, S7, S30, S31, S2 and S17 are the top-10 subjects. The $P_{acc}$ of top-10 subjects achieves 77.18±4.12%, and the corresponding $P_f$ is 75.52±4.24%. Overall, the proposed method has achieved the outstanding performance.

TABLE I
The results of the proposed method in the subject-independent experiment (%)

| Strategies | LA vs. HA | | LV vs. HV | |
|---|---|---|---|---|
| | $P_{acc}$ | $P_f$ | $P_{acc}$ | $P_f$ |
| All subjects | 68.13 (8.45) | 66.61 (8.53) | 67.03 (8.08) | 65.87 (7.87) |
| Top-10 subjects | 78.68 (4.88) | 77.16 (4.68) | 77.18 (4.12) | 75.52 (4.24) |

Note: The standard deviations are listed in the brackets.

### C. Comparison with related works

We compare our method with other cross-subject methods on the DEAP dataset. The performances of different methods are listed in the Table II. Overall, the proposed method has achieved the best accuracy on the arousal and valence classification. Only the $P_f$ of valence classification is less than the Ref. [32] by 1.02%.

TABLE II
The related works on DEAP dataset using cross-subject strategy (%)

| Method | LA vs. HA | | LV vs. HV | |
|---|---|---|---|---|
| | $P_{acc}$ | $P_f$ | $P_{acc}$ | $P_f$ |
| SSFE&SVM [31] | 65.21 | 62.49 | 66.35 | 64.52 |
| LRFS [32] | 67.97 | 62.99 | 65.10 | **66.89** |
| 1D-CNN [20] | 63.75 | 63.35 | 62.27 | 65.37 |
| CapsNet [33] | 58.53 | - | 48.29 | - |
| AP-CapsNet [34] | 63.51 | - | 62.71 | - |
| TCN&ADDA [8] | 64.33 | - | 63.25 | - |
| BiVDANN [10] | - | - | 63.52 | - |
| HSLT [15] | 65.75 | 64.29 | 66.51 | 66.27 |
| Proposed method | **68.13** | **66.61** | **67.03** | 65.87 |

Note: SSFE denotes Shared-subspace feature elimination; SVM denotes support vector machine; LRFS denotes Locally-Robust Feature Selection; CapsNet denotes Capsule neural networks; AP-CapsNet denotes attention mechanism and the pre-trained convolutional CapsNet network; BiVDANN denotes Bilateral Variational Domain Adversarial Neural Network.

In the Ref. [32], a feature selection method is proposed to select the features with high consistency among all the subjects. Different from LRFS, the proposed method could adaptively learn domain-invariant features by DANN. And the HSLT [15] is a transformer-based spatial learning method in our previous work, and handcrafted PSD features are fed to the transformer. Compared with the HSLT, the end-to-end temporal-spatial feature learning teacher model is an improved version of HSLT, and the goal of this work is to enhance the domain-invariant feature learning rather than just focus on spatial feature learning. In the Ref. [8], TCN and ADDA are combined for improving the cross-subject EEG emotion recognition. TCN has the strong ability to learn temporal context, but spatial correlation among the electrodes could not be well captured when compared with TSFIN. In the Ref. [10], bi-hemisphere asymmetry learning and DANN are combined to learn the domain-invariant features. Different from the proposed method, the variational autoencoders, which are trained by labeled and unlabeled samples, are adopted to deal with the limited EEG data problems.

Next, we compare our model with related works in terms of performance and number of parameters in Table III. Although EEGNet, DenseNet and CNN-LSTM have fewer parameters, the proposed method outperforms these methods more than 5%. Besides, our model has fewer parameters than HSLT and the teacher model. The proposed method outperforms the HSLT by 2.38% on arousal classification and 0.52% on valence classification, and the proposed method is slightly outperformed by the teacher model. Moreover, the number of the parameters within our model is only about 22% of the teacher model and 55% of the HSLT. And it also should be noted that the proposed method is a two-stage method, so it consumes more computation.

Furthermore, the limited sample quantity might be the reason that the transformer model could not significantly outperform the proposed method. And we also find that the performance would be decreased when transformer models without DANN, this might be due to the strong individual differences of EEG. On the other hand, the transformer models still obviously outperform the conventional deep learning model. The reason might be that the multi-head self-attention mechanism has the strong ability of sequence learning. Hence, we adopt the transformer model to assist the student model to learn complex temporal and spatial information. In general, the proposed method achieves outstanding cross-subject performance with fewer parameters.

TABLE III
The performance and parameter comparison with related works on DEAP dataset using cross-subject strategy (%)

| Method | LA vs. HA | | LV vs. HV | | Params |
|---|---|---|---|---|---|
| | $P_{acc}$ | $P_f$ | $P_{acc}$ | $P_f$ | |
| CNN-LSTM* [35] | 62.31 | 60.70 | 62.74 | 60.15 | 0.63M |
| EEGNet* [36] | 61.93 | 59.14 | 61.32 | 59.57 | 0.05M |
| HSLT [15] | 65.75 | 64.29 | 66.51 | 66.27 | 1.22M |
| DenseNet* [37] | 62.03 | 61.39 | 60.96 | 59.48 | 0.19M |
| Teacher model | **68.87** | **66.76** | **67.59** | **66.44** | 3.02M |
| Proposed method | 68.13 | 66.61 | 67.03 | 65.87 | 0.67M |

Note: The model with * denotes that we conduct same experiment by our own reimplementation.

## D. Comparison with related works

To validate the effectiveness of the two steps and feature interaction module, we conduct the ablation study and the experimental results are shown in the Table. IV. The performance of the proposed method significantly outperforms other strategies. The lowest $P_{acc}$ has been obtained when the KD stage and DANN stage are both removed. This could validate the effectiveness of the combination of KD and DANN. We also find that the performance of the proposed method has significantly decreased (more than 4%) when the DANN stage is removed. It could validate that domain-invariant feature learning is the essential part of the proposed method. Besides, the $P_{acc}$ degrades even more when KD stage is removed (more than 5%). It indicates that it is essential to assist the student model to learn temporal-spatial representation within KD stage. The reason might be that we design the teacher model and student model with similar structure, and the meanings of the transfer latent features are the same. Our strategy is basically consistent with the finding in the [11, 18] that the similarity between the student and teacher is highly related to how well the student can mimic the teacher. Finally, although $P_{acc}$ slightly degrade when feature interaction module is removed, this lightweight module still makes contributions to the overall performance.

TABLE IV
The results of the ablation study (%)

| Strategy | LA vs. HA | | LV vs. HV | |
|---|---|---|---|---|
| | $P_{acc}$ | $P_f$ | $P_{acc}$ | $P_f$ |
| Proposed method | **68.13*** | **66.61** | **67.03*** | **65.87** |
| w/o K stage | 62.76 | 60.78 | 61.87 | 59.91 |
| w/o D stage | 64.30 | 62.88 | 63.45 | 61.39 |
| w/o K&D stages | 61.85 | 60.39 | 60.26 | 58.76 |
| w/o FIM | 67.27 | 65.33 | 65.89 | 63.96 |

*Note: FIM denotes feature interaction module; K stage denotes KD stage; D stage denotes DANN stage. The accuracy with * denotes that the proposed method significantly outperforms other strategies using paired t-test (p<0.05).*

## V. DISCUSSION

To further verify the effectiveness of the proposed method, we conduct feature visualization and ablation study of teacher model. And we discuss the advantages and limitations of the proposed method.

### A. The feature visualization of the proposed method

To further validate the effectiveness of the proposed method, we visualize the raw data and feature aggregation of the TSFIN using t-SNE algorithm [38]. In the Table V and Table VI, we can find that it exists higher differences between the raw data distributions of source domain and target domain. It indicates the difficulty of the cross-subject emotion recognition due to strong individual differences among the subjects. In the distribution of feature aggregation, we can find that the target domain and source domain are both generally divided into two clusters (the high class and low class), and decision boundary of target domain within S13 is clear, S14 and S15 are slightly inferior. Moreover, the feature distribution of target domain can both be well aligned with that of source domain within two clusters. It demonstrates that the domain-alignment have been achieved by proposed method.

TABLE V
The feature visualization of arousal classification

| Sub. No. | Raw features | Feature Aggregation of TSFIN |
|---|---|---|
| S13 | 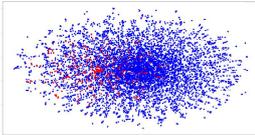 | 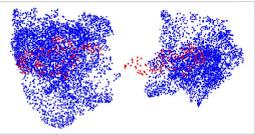 |
| S14 | 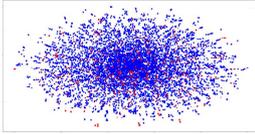 | 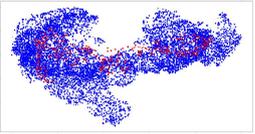 |
| S15 | 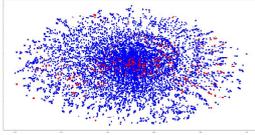 | 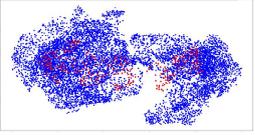 |

*Note: Sub. No. denotes the number of the subject.*

TABLE VI
The feature visualization of valence classification

| Sub. No. | Raw features | Feature Aggregation of TSFIN |
|---|---|---|
| S13 | 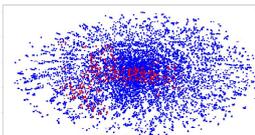 | 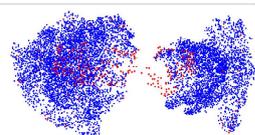 |
| S14 | 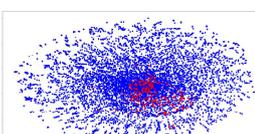 | 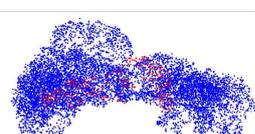 |
| S15 | 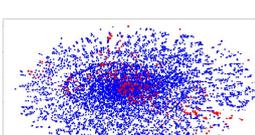 | 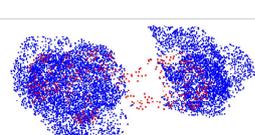 |

Besides, the performance of domain-alignment and decision boundary of target domain within valence classification is lower than that of arousal classification. This might verify that the $P_{acc}$ of arousal classification is higher than the $P_{acc}$ of valence classification. To sum up, these results could validate the effectiveness of the proposed method and demonstrate the difficulty of cross-subject EEG classification.

### B. The Ablation study of teacher model

The design of the teacher model is also essential to the proposed method. Hence, we conduct the ablation study of the teacher model. In the Table VII, the proposed teacher model has achieved the best performances among all the models. It achieves the accuracy of 68.87% and 67.59% at arousal and valence level. Compared with the strategy of without temporal feature extractor and strategy of without hierarchical spatial learning, TSERT have boosted the accuracy more than 3%. This could validate the effectiveness of temporal feature extraction module and hierarchical spatial learning module. Besides, we also compare the impact of different input forms on performance of the proposed teacher model. According to the results, we find that the performance of raw EEG signals surpasses PSD features. It indicates that the proposed teacher model could take full advantage of the detailed dynamic temporal information which is beneficial to discriminate the emotions.

TABLE VII
The ablation study results of the proposed teacher model

| Strategy | LA vs. HA | | LV vs. HV | |
|---|---|---|---|---|
| | Acc | F1 | Acc | F1 |
| Proposed teacher model | **68.87** | **66.76** | **67.59** | **66.44** |
| w/o TFE | 64.98 | 63.31 | 64.48 | 62.11 |
| w/o HSL | 65.02 | 63.56 | 63.71 | 61.98 |
| w/ PSD input | 66.76 | 64.42 | 65.73 | 63.51 |

Notes: TFE denotes temporal feature extractor; HSL denotes hierarchical Spatial learning; w/ PSD input denotes the proposed teacher model with PSD input.

### C. The advantages and limitations of the proposed method

In this work, we elaborately design the KD stage to help student model, which is guided by a transformer model, to learn complex temporal dynamics and spatial correlation of EEG. Specifically, we take full advantage of the ability of the transformer model to learn the sequence context, and the problem of the cumbersome feature extractor in the DANN could be alleviated by this strategy. Besides, temporal-spatial feature interaction, which effectively improve the performance, is utilized throughout the two stages. This strategy is on the basis of multi-representation adaptation network [17] (Representing the features from different aspects) and our previous work [13] (Temporal and spatial feature fusion).

On the other hand, the limitations of this work also should be discussed. The proposed method has achieved a better performance with fewer parameters, but it consumes more computation than the single stage method. Besides, we aim to validate the effectiveness of combining KD with DANN in this work. The DANN stage still has the space for the improvement. We assume that the subjects in the source domain are independent identically distributed. In fact, multi-source transfer learning should be adopted to close to the practice.

## VI. EXPERIMENTAL RESULTS

In this work, we propose a lightweight DANN based on KD to improve the performance of the cross-subject EEG emotion recognition. The effectiveness of proposed method has been verified by the sufficient experiment results. In the KD stage, we adopt a transformer model to guide the student model to learn the complex temporal and spatial information of EEG. And the lightweight student model with a feature interaction module is served as the feature extractor within DANN stage. In the proposed method, we take full advantage of the strong sequence learning ability of the transformer model, and robust lightweight model is more suitable to the DANN. Besides, the temporal-spatial feature interaction within two stages is also contributive to the performance. In the future work, we will adopt multi-source transfer learning method to further improve the proposed method.